\documentclass[a4paper,12pt]{article}
\usepackage[dvips]{graphics}
\newcommand{\PPEnum}    {CERN-EP/98-089}

\newcommand{\Date}    {29 May 1998}

\newcommand{\definmath}[2] {\def#1{\ifmmode#2\else$#2$\fi}}
\newcommand{\deriv}[1]     {{\mathrm{d}}#1}
\def\etal{{\sl et al.}}
\def\epem {e$^{+}$e$^{-}$} 
\def\z  {\mbox{${\rm Z}^{0}$}} 
\def\gev{\mbox{\rm GeV}}
\def\gevc{\mbox{\rm GeV/$c$}}
\def\gevcc{\mbox{\rm GeV/$c^2$}}
\def\xmat{\mbox{${\cal M}$}}
\def\fr{\mbox{${w}$}}

\def\tagcorr{\mbox{${\cal T}$}}

\def\Dstar{\mbox{${\rm D}^{*\pm}$}}
\def\Dstarp{\mbox{${\rm D}^{*+}$}}
\def\smallDstar{{\mbox{\scriptsize ${\rm D}^{*\pm}$}}}
\def\D0{\mbox{${\rm D}^{0}$}}
\def\smallD0{ \mbox{{\scriptsize ${\rm D}^{0}$}}}
\def\declsig{\mbox{$L/{\sigma_L}$\ }}

\def\xe{\mbox{$x_{p}$}}
\def\xemean{\mbox{$ \langle x_{p}\rangle$}}
\def\xip{\mbox{$\xi_{p}$}}
\def\ximax{\mbox{$\xi_{0}$}}
\definmath{\xE} {x_p}
\def\xdstar{\mbox{$x_{D^{*}}$}}
\def\declsig{\mbox{$L/\sigma_{L}$\ }}

\definmath{\roots} {\sqrt{s}}
\definmath{\Ecm} {E_{\mathrm{cm}}}
\definmath{\Ebeam}  {E_{\mathrm{b}}}
\definmath{\as} {\alpha_s}
\definmath{\Evis}   {E_{\mathrm{vis}}}
\newcommand{\PhysLett}  {Phys.~Lett.}
\newcommand{\PRL} {Phys.~Rev.\ Lett.}

\newcommand{\PhysRev}   {Phys.~Rev.}
\newcommand{\NPhys}  {Nucl.~Phys.}
\newcommand{\NIM} {Nucl.~Instr.\ Meth.}
\newcommand{\CPC} {Comp.~Phys.\ Comm.}
\newcommand{\ZPhys}  {Z.~Phys.}

\definmath{\PZz} {\mathrm{Z}^{0}}      
\definmath{\Pq}      {q}
\definmath{\Paq}  {\overline{q}}
\begin{document}
%

 
\begin{titlepage}
%

\begin{center}
    \large
    EUROPEAN LABORATORY FOR PARTICLE PHYSICS
\end{center}

\begin{flushright}
    \large
    \PPEnum\\
    \Date
\end{flushright}

%
%
\bigskip
\begin{center}
    \huge\bf\boldmath
    Measurements of Flavour Dependent Fragmentation Functions in
    ${\bf Z^{0} \rightarrow {\rm q}\bar{\rm q}}$ Events
\end{center}
%
%
\begin{center}
\Large
The OPAL Collaboration \\
\bigskip
\end{center}
\bigskip
%
%
\begin{abstract}
Fragmentation functions for charged particles in $\z \rightarrow 
{\rm q}\bar {\rm q}$
events have been measured for bottom (b), charm (c) and light (uds) quarks  as well 
as for all flavours together. 
The results are based on data recorded between 1990 and 1995 using the OPAL
detector at LEP. Event samples with different flavour compositions were formed using
reconstructed \Dstar \,mesons and
secondary vertices. The $\xip =\ln(1/\xe)$ distributions and the position of their
maxima \ximax\ are also presented separately for uds, c and b
quark events. The fragmentation function for b quarks is significantly
softer than for uds quarks.

\end{abstract}
 
\smallskip
 
\smallskip
\begin{center}
\end{center}
 
\smallskip
\begin{center}
{\large 
Submitted to European Physics Journal C }\\
\end{center}
 
\end{titlepage} \
%
%
\begin{center}{\Large        The OPAL Collaboration
}\end{center}\bigskip
\begin{center}{
K.\thinspace Ackerstaff$^{  8}$,
G.\thinspace Alexander$^{ 23}$,
J.\thinspace Allison$^{ 16}$,
N.\thinspace Altekamp$^{  5}$,
K.J.\thinspace Anderson$^{  9}$,
S.\thinspace Anderson$^{ 12}$,
S.\thinspace Arcelli$^{  2}$,
S.\thinspace Asai$^{ 24}$,
S.F.\thinspace Ashby$^{  1}$,
D.\thinspace Axen$^{ 29}$
% ...
}\end{center}\bigskip
\bigskip
$^{  1}$School of Physics and Astronomy, University of Birmingham,
Birmingham B15 2TT, UK
\newline
$^{  2}$Dipartimento di Fisica dell' Universit\`a di Bologna and INFN,
I-40126 Bologna, Italy
\newline
$^{  3}$Physikalisches Institut, Universit\"at Bonn,
D-53115 Bonn, Germany
\newline
$^{  4}$Department of Physics, University of California,
Riverside CA 92521, USA
\newline
$^{  5}$Cavendish Laboratory, Cambridge CB3 0HE, UK
\newline
$^{  6}$Ottawa-Carleton Institute for Physics,
Department of Physics, Carleton University,
Ottawa, Ontario K1S 5B6, Canada
\newline
$^{  7}$Centre for Research in Particle Physics,
Carleton University, Ottawa, Ontario K1S 5B6, Canada
\newline
$^{  8}$CERN, European Organisation for Particle Physics,
CH-1211 Geneva 23, Switzerland
\newline
$^{  9}$Enrico Fermi Institute and Department of Physics,
University of Chicago, Chicago IL 60637, USA
\newline
$^{ 10}$Fakult\"at f\"ur Physik, Albert Ludwigs Universit\"at,
D-79104 Freiburg, Germany
\newline
$^{ 11}$Physikalisches Institut, Universit\"at
Heidelberg, D-69120 Heidelberg, Germany
\newline
$^{ 12}$Indiana University, Department of Physics,
Swain Hall West 117, Bloomington IN 47405, USA
\newline
$^{ 13}$Queen Mary and Westfield College, University of London,
London E1 4NS, UK
\newline
$^{ 14}$Technische Hochschule Aachen, III Physikalisches Institut,
Sommerfeldstrasse 26-28, D-52056 Aachen, Germany
\newline
$^{ 15}$University College London, London WC1E 6BT, UK
\newline
$^{ 16}$Department of Physics, Schuster Laboratory, The University,
Manchester M13 9PL, UK
\newline
$^{ 17}$Department of Physics, University of Maryland,
College Park, MD 20742, USA
\newline
$^{ 18}$Laboratoire de Physique Nucl\'eaire, Universit\'e de Montr\'eal,
Montr\'eal, Quebec H3C 3J7, Canada
\newline
$^{ 19}$University of Oregon, Department of Physics, Eugene
OR 97403, USA
\newline
$^{ 20}$Rutherford Appleton Laboratory, Chilton,
Didcot, Oxfordshire OX11 0QX, UK
\newline
$^{ 22}$Department of Physics, Technion-Israel Institute of
Technology, Haifa 32000, Israel
\newline
$^{ 23}$Department of Physics and Astronomy, Tel Aviv University,
Tel Aviv 69978, Israel
\newline
$^{ 24}$International Centre for Elementary Particle Physics and
Department of Physics, University of Tokyo, Tokyo 113, and
Kobe University, Kobe 657, Japan
\newline
$^{ 25}$Institute of Physical and Environmental Sciences,
Brunel University, Uxbridge, Middlesex UB8 3PH, UK
\newline
$^{ 26}$Particle Physics Department, Weizmann Institute of Science,
Rehovot 76100, Israel
\newline
$^{ 27}$Universit\"at Hamburg/DESY, II Institut f\"ur Experimental
Physik, Notkestrasse 85, D-22607 Hamburg, Germany
\newline
$^{ 28}$University of Victoria, Department of Physics, P O Box 3055,
Victoria BC V8W 3P6, Canada
\newline
$^{ 29}$University of British Columbia, Department of Physics,
Vancouver BC V6T 1Z1, Canada
\newline
$^{ 30}$University of Alberta,  Department of Physics,
Edmonton AB T6G 2J1, Canada
\newline
$^{ 31}$Duke University, Dept of Physics,
Durham, NC 27708-0305, USA
\newline
$^{ 32}$Research Institute for Particle and Nuclear Physics,
H-1525 Budapest, P O  Box 49, Hungary
\newline
$^{ 33}$Institute of Nuclear Research,
H-4001 Debrecen, P O  Box 51, Hungary
\newline
$^{ 34}$Ludwigs-Maximilians-Universit\"at M\"unchen,
Sektion Physik, Am Coulombwall 1, D-85748 Garching, Germany
\newline
\bigskip\newline
$^{  a}$ and at TRIUMF, Vancouver, Canada V6T 2A3
\newline
$^{  b}$ and Royal Society University Research Fellow
\newline
$^{  c}$ and Institute of Nuclear Research, Debrecen, Hungary
\newline
$^{  d}$ and Department of Experimental Physics, Lajos Kossuth
University, Debrecen, Hungary
\newline
$^{  e}$ on leave of absence from the University of Freiburg
\newline
\newpage
%
\section{Introduction}
%
Experimental measurements of the inclusive momentum 
distribution of charged particles 
in \epem \ collisions provide 
important insight into the process of how quarks turn into 
hadrons. This  distribution is commonly normalised to the total hadronic cross 
section $\sigma^{\mathrm{tot}}$ and presented as a function of the scaled momenta
$\xE = 2 p_h/\roots$ of the charged hadrons, 
where \roots\ is the centre-of-mass energy.
 In this form, the spectrum is
usually referred to as the fragmentation function and can be obtained experimentally
from the total number of hadronic final states, $N_{\rm event}$, and the
 number of charged
particles in each \xe\ bin, $N_{\rm track}(\xe)$:
\begin{equation}
\label{eqn:1}
  F(\xE) = 
      \frac{1}{\sigma^{\mathrm{tot}}}\frac{\deriv{\sigma^h}}{\deriv{\xE}}= \frac{1}{
  N_{\rm event}}\frac{N_{\rm track}(\xe)}{\Delta \xE}.
\end{equation}
The charged particle momentum spectrum can also be studied as the 
 distribution of $\xip =\ln(1/\xe)$. The \xip\ distribution
emphasises the low momentum component and
the \xe \ distribution  the high momentum component of the momentum spectrum.

In the na\"{\i}ve quark parton model,  the scaled momentum distribution is expected
to be independent from the centre-of-mass energy. A violation
of this scaling is expected due to gluon radiation in the final state.
Experimentally, scaling violation in fragmentation
functions had indeed been observed  by 
combining measurements at different centre-of-mass energies
and could be used to determine
\as\ \cite{delphi1}. The position of the maximum of the \xip\ distribution, \ximax,
has been studied in the past in various experiments 
(see for example \cite{highE}
and references therein). The energy dependence of the position of the maximum
provides an important test of the QCD prediction for the emission of soft 
gluons \cite{xi_nlla}.

In events with a heavy primary quark, the possibility of cascade decays of bottom or
charm hadrons results in more particles sharing the same energy than
in light quark events and a softer momentum spectrum can be expected.
Since the flavour composition of the primary quarks in 
${\rm e}^{+}{\rm e}^{-} \rightarrow {\rm q}\bar{\rm q}$ is predicted by the
electroweak theory to change with centre-of-mass energy, this flavour dependence 
of the momentum spectra affects the energy dependence of the \xe\ and \xip\ 
distributions of the inclusive event sample.
To correct for this contribution, in \cite{aleph} not only 
the inclusive fragmentation function was measured but also fragmentation
functions in event samples 
with different flavour compositions were studied. 
In \cite{delphi2}, measurements of fragmentation functions in samples with different 
flavour composition were used to extract flavour dependent fragmentation
functions for events with 
primary light (uds),  charm (c) or bottom (b) quarks.

Here we present a 
measurement of flavour dependent fragmentation functions, based on 
the methods developed for the OPAL measurement of
charged particle multiplicities in uds, c and b quark events 
\cite{opalb,opalc}. 
Events were divided into two hemispheres by a plane perpendicular
to the thrust axis. Secondary vertices and reconstructed \Dstar \,mesons were
used to tag
hemispheres to create samples of events with different quark flavour 
mixtures (Section 2). 
To reduce the biases induced by the tagging, the measurement of the 
fragmentation functions was based on the 
momentum spectrum of charged particles 
in the event hemisphere opposite to the tag. Corrections for hemisphere 
correlations and for 
distortions due to detector effects are described in Section 3. The flavour 
dependent \xe\ and \xip\ distributions were obtained  from a simultaneous fit 
to the momentum spectra
of the different hemisphere samples (Section 4).
For the first time, the measured position of the maximum of 
the \xip\ distribution, \ximax, is presented separately for
uds, c and b events. 
A measurement of the inclusive distribution of all five flavours 
was also performed, based on the track
momentum spectrum of all events, i.e.,~without considering any flavour tagging.
%
%
\section{Selection and event tagging}
\label{l_Selcection}
%
%
A complete description of the OPAL detector can be found elsewhere
\cite{opaldet}. This analysis relied on the precise reconstruction of charged
particles in the central detector, consisting of a silicon microvertex
detector, a vertex drift chamber, a large volume jet chamber and
chambers measuring the $z$-coordinate%
\footnote{The OPAL coordinate system is
defined with positive $z$ along the electron beam direction and with positive
$x$ pointing towards the centre of the LEP ring. The polar angle
$\theta$ is defined relative to the $+z$ axis and the azimuthal angle $\phi$
relative to the $+x$ axis.}
of tracks as they leave the jet chamber.

This analysis used data recorded with the OPAL detector
 in the years 1990 to 1995 at centre-of-mass energies around 91.2 GeV comprising
 an integrated luminosity of about 177~pb$^{-1}$.
  \z\ decays were selected using the criteria described  
  in \cite{opalmh}. 
To ensure that most charged particles were well contained in the detector,
the polar angle of the thrust axis was required to satisfy
$|\cos \theta_{\rm thrust}| < 0.8$. 
To reduce systematic errors in the application of the secondary vertex tag, it
was only based on a homogeneous data sample taken in the year 1994,
 representing an integrated luminosity of about 34 pb$^{-1}$.
The full integrated luminosity was used in the case of the 
\Dstar~meson tag.

Charged tracks used in the measurement 
of the fragmentation function were required to have a measured momentum 
in the \mbox{$x$-$y$} plane, $p_t$, of at least 0.150
GeV/$c$ and to satisfy $|d_0| < 0.5$ cm, where $d_0$ is the distance of
closest approach to the origin in the $x$-$y$ plane. 

Simulated hadronic \z\, decays were generated with the Jetset 7.4
 Monte Carlo program \cite{jetset} tuned to OPAL data \cite{opaltune}. The
 events were
passed through a detailed simulation of the OPAL detector \cite{opalsim} and
processed using the same reconstruction and selection algorithms as the data.

\subsection{Secondary vertex tag}
Samples with varying purity of b quark events were selected by reconstructing
secondary vertices, following the procedure described in \cite{opalb}. 
Events were divided into two hemispheres
by the plane perpendicular to the thrust axis and comprising the interaction point.
 Jets were
reconstructed by combining charged tracks and electromagnetic clusters 
not associated to tracks, using the scaled invariant mass
algorithm described in \cite{opale0} with the JADE-E0 recombination scheme 
and the invariant
mass cut-off being set to $7\,\gevcc$. 
A vertex fit was then attempted in the highest energy jet in each hemisphere
separately.
Each track used in these vertex fits was required to have at least one hit in the
silicon microvertex detector. All such tracks in the jet were fitted to a common 
vertex
point in the $x$-$y$ plane and the track with the largest contribution to the
$\chi^2$ was removed if this contribution was greater than four. 
The remaining tracks were then refitted until either all tracks contributed less than
four to the $\chi^2$ 
 or there were fewer than four remaining tracks. For each
successfully reconstructed secondary vertex, the projected decay length $L$  
in the $x$-$y$ plane with respect to the primary vertex was calculated, where 
the primary vertex was reconstructed from all tracks in the event
together with a constraint to the average beamspot position as in \cite{primvert}.

\begin{figure}[htbp]
\begin{center}
\resizebox{160mm}{!}{\includegraphics*{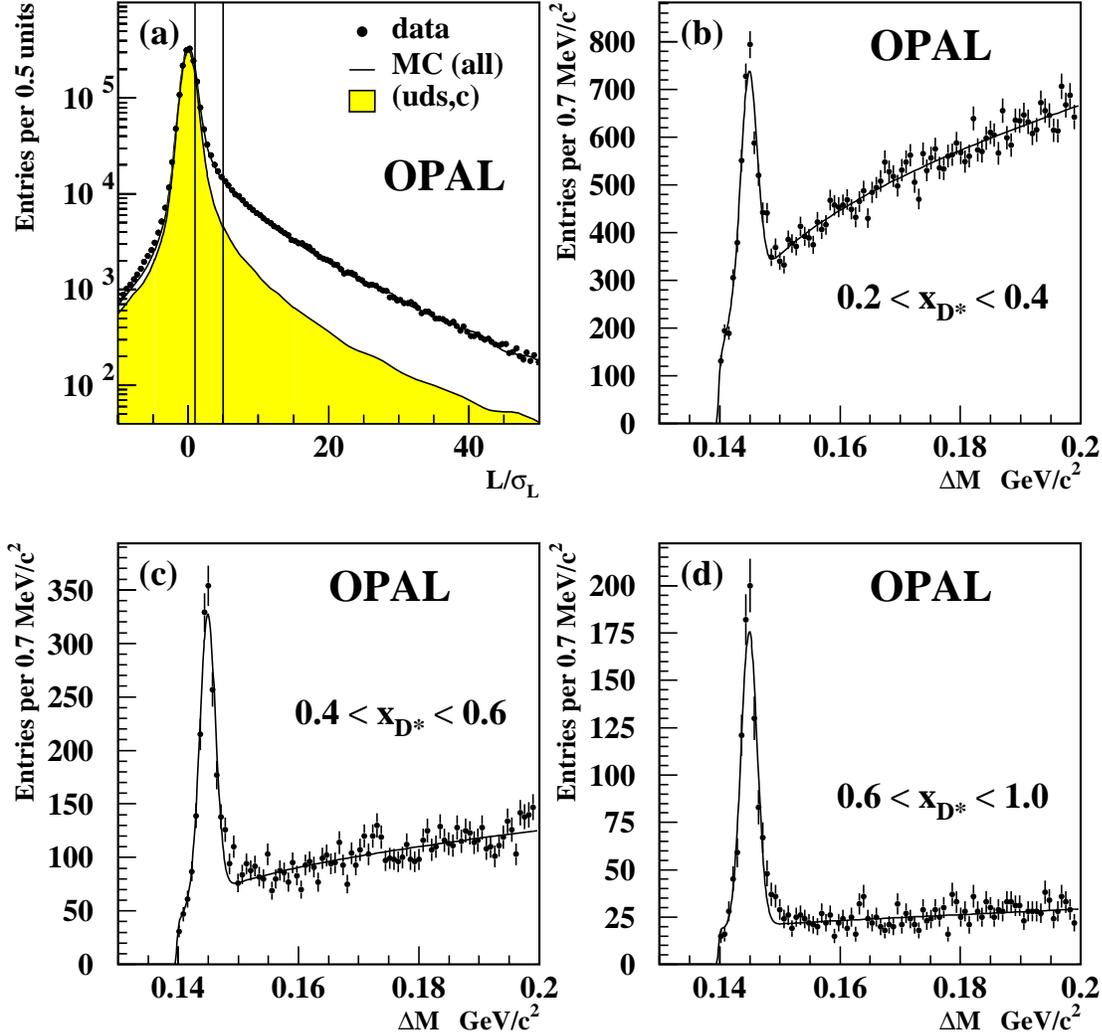}}
\end{center}
\caption[ ]{\sl (a): The decay length significance distribution in
 data (symbols) and 
 Monte Carlo (solid curve). The contribution from uds and
 from c quarks in the Monte Carlo distribution has been shaded.
  The boundaries of the three decay length 
  significance bins
 used in this analysis: $-10<\declsig<1$, $1<\declsig<5$ and 
 $5<\declsig<50$ are indicated by vertical lines. (b) to (d): The distribution 
 of the mass difference of the \Dstar\,candidate and \D0\,candidate in the three 
different \xdstar \ bins. The symbols show the data while the solid lines 
are the results of the fits described in the text.
\label{plot8}}
\end{figure}

The decay length significance, i.e.~the decay length divided by its uncertainty
 \declsig\ was used to obtain three event 
samples $k$ of varying b flavour purity. According to the simulation, the b purities 
$f_k^{\rm b}$ 
in these samples vary from $11 \%$ to $90 \%$ (see Table \ref{Table1}). 
Fig \ref{plot8}(a) shows the distribution of the decay length significance 
in data and in Monte Carlo.
%
%
\begin{table}[hbt]
\begin{center}
\begin{tabular}{|c|c|c|c|}  
                            \hline 
 & $-10.0 < \declsig  < 1.0 $&$ 1.0 < \declsig < 5.0 $&$ 5.0 < \declsig < 50.0 $\\  
                            \hline
     Number of hemispheres & 940\,275   & 268\,500  & 117\,665  \\
     uds quark fraction $f_k^{\rm uds}$       &   0.71     &    0.39   &    0.04   \\
     c quark fraction $f_k^{\rm c}$         &   0.18     &    0.23   &    0.06   \\
     b quark fraction $f_k^{\rm b}$         &   0.11     &    0.38   &    0.90   \\
                                           \hline
\end{tabular}
\caption {\sl Number of tagged hemispheres in data 
and the flavour composition derived from the Monte Carlo simulation 
in three decay length significance regions. \label{Table1}}
\end{center}
\end{table} 
%
%

%
\subsection{\boldmath \Dstar \,meson tag}
Event samples with an enriched c quark contribution were obtained by
reconstructing \Dstar\,me\-son candidates.
\Dstar\,candidates were selected via the decay%
\footnote{Throughout this paper, charge conjugate particles and decay
 modes are always implied.}
$\Dstarp \rightarrow
{\rm K}^-\pi^+\pi^+$ closely following the procedure described in \cite{opalc}:
\begin{itemize}

\item A subset of tracks was selected that have 
 $p_t > 0.250\  \gevc$ and $|d_0| < 0.5$ cm.

\item Candidates of $\D0 \rightarrow
{\rm K}^-\pi^+$ decays were selected by taking all combinations of two
oppositely charged tracks, with one of them assumed to be a pion and the other
assumed to be a kaon. \Dstar \,can\-di\-dates were selected by combining \D0
can\-di\-dates with a third track. This `{}slow pion{}' track was required to have
the same charge as the track presumed to be the pion in the \D0 decay.

\item The probability  that the measured rate 
of energy loss, ${\rm d}E/{\rm d}x$, for the kaon candidate track was
 consistent with that expected for
a real kaon was required to be greater than $10 \%$. 

\item At least two of the three tracks were required to have either $z$-chamber 
hits or a polar angle measurement derived from the point at which the track has left 
the jet chamber. 

\item The invariant mass of the \D0\,candidate was required to be between
1.790\,\gevcc\ and 1.940\,\gevcc\ and the mass difference between the \D0\,and the
\Dstar\,candidate, $\Delta M$, was required to be between 0.142\,\gevcc\ and
0.149\,\gevcc.

\item Making use of the fact that real \D0 mesons decay isotropically in their rest
frames whereas combinatorial background is peaked in 
the forward and the backward direction, the
following cuts were applied: $|\cos{\theta^*}|<0.8 $ for $ \xdstar < 0.5 $ and
  $|\cos{\theta^*}|<0.9 $ for $  \xdstar >0.5 $,
where $\theta^*$ is the angle between the kaon in the \D0 rest frame and the
direction of the \D0 in the laboratory frame and \xdstar \ is the scaled energy of
the \Dstar, i.e., $\xdstar = 2E_{\smallDstar} / \roots $.

\end{itemize}
To provide samples with differing charm purity, the data were divided into
three  \xdstar \ regions.
To evaluate their flavour composition, $\Delta M$  distributions obtained 
without the
cut on $\Delta M$ were
fitted with a Gaussian for the signal and  a function
$A\exp(-B\Delta M)(\Delta M/m_{\pi} - 1 )^C$ for the background
\cite{opaldstar}. The signals, together with the fitted functions
 are shown in Fig.\,\ref{plot8}(b)-(d).
These fits were used to determine the fraction of background,
$f^{\rm BG}_k$ in each \Dstar \,sample in the signal $\Delta M$ region.
 The results of these fits are summarised in
Table \ref{Table_dstar}.

%
%
\begin{table}[hbt]
\begin{center}
\begin{tabular}{|c|c|c|c|}   
                            \hline
     &  $0.2 < \xdstar < 0.4$ &$0.4 < \xdstar < 0.6$ &$0.6 < \xdstar < 1.0$ \\
                            \hline
Number of \Dstar \,candidates         &  5109   &  1951 & 985 \\ 
combinatorial background fraction $f^{\rm BG}_k$    &  0.57   & 0.36  & 0.24  \\
c quark fraction ${\cal P}^{\rm c}_k$&$0.22\pm 0.06$&$0.50 \pm 0.06$&$0.90\pm 0.04 $\\
b quark fraction ${\cal P}^{\rm b}_k$&$0.78\pm 0.06$&$0.50 \pm 0.06$&$0.10\pm 0.04 $\\
                                           \hline
\end{tabular}
\caption[ ] {\sl Number of  \Dstar \,candidates, the fitted background fraction 
and
 the flavour composition of events with a genuine \Dstar\,as taken from 
 \cite{opaldstar} in three \xdstar\ regions.
\label{Table_dstar}}
\end{center}
\end{table} 
%
%
The selected samples of \Dstar\,candidates have three components: genuine 
\Dstar\,mesons from b quark decays, genuine \Dstar\,mesons from c quark decays
 and combinatorial background.
No other sources of \Dstar\,candidates were considered since Monte Carlo 
simulations predicts that
only $0.3\%$ of \Dstar\,mesons with $\xdstar>0.2$ are produced via 
gluon splitting in light quark events \cite{opaldstar}.
To evaluate the effect of the contribution from fake \Dstar,
a side-band sample was selected by requiring that the two pions of the 
\Dstar \,candidates had opposite charge
and that $0.150\ \gevcc < \Delta M \ <  0.170\ \gevcc$. Once this contribution was
taken into account, the flavour composition of the \Dstar~samples was 
taken as the fractions ${\cal P}_k^{\rm c}$ and 
${\cal P}_k^{\rm b}$ of genuine
\Dstar~mesons originating from a primary c quark and b quark as measured
 in \cite{opaldstar}. 
While in \cite{opaldstar} the fractions ${\cal P}_k^{\rm c}$ and 
${\cal P}_k^{\rm b}$ were 
derived for \Dstar\,candidates after corrections  for detector 
efficiency and acceptance were made, they were applied in this analysis to 
uncorrected 
data. No modifications were made since tests with Monte Carlo simulated events
showed no significant flavour dependence of these corrections. 
%
\section{Corrections}

The track momentum distributions in the hemispheres opposite to the secondary 
vertex 
or \Dstar\ tag were measured. Six distributions (label $k$) were obtained, 
corresponding to the three decay length regions and the three
\xdstar\ regions. To obtain fragmentation functions from these
distributions, three sets of corrections were applied. Firstly, a correction 
was made to
take into account track momentum resolution and 
reconstruction efficiency.  Secondly, the effects due to 
the event selection and the correlation between  hemispheres were 
accounted for.
In addition, the measured track momentum spectra in the \Dstar\ tag samples were
corrected for the contribution of fake \Dstar\ mesons. 
The different corrections are described in the following. 

After this procedure, the track momentum distributions are defined as the momentum
distributions of all promptly produced stable charged particles and those produced
in the decays of particles with lifetimes shorter than $3 \times 10^{-10}$ sec.,
corrected for initial state radiation.
This means that charged decay products from ${\rm K}_{\rm s}^0$, hyperons and
weakly decaying b and c flavoured hadrons are included in the definition,
regardless of how far away from the interaction point the decay actually occurred.

\subsection{Track momentum resolution and efficiency}
\label{corr.track}

The number $N_{j,k}^{\rm observed}$ of  tracks in a tag sample
was measured in 22 different \xe\ bins
$j$.\footnote{ For the measurement of the \xip\ distribution, a different binning
   with 29 \xip\ bins was used. Apart from the binning, there were no
   differences between the analysis of the \xe\
   and the \xip\ distribution, so the measurement of the
   \xip\ distribution is not explicitly described in the following sections.}
The corrected distribution for a given tag sample is
\begin{equation}
{N}^{\rm corrected}_{i,k} = 
\sum_j \sum_{q} 
\frac{\xmat_{ij}^{q}}{\epsilon_{i}^q}(\fr_{j,k}^q N^{\rm observed}_{j,k}).
\end{equation}
Here, $\xmat_{ij}^{q}$ is the probability that a 
track measured in  \xe \ bin $j$ originates from a true \xe\ bin $i$.
This correction was applied to account for the migration of the tracks
between different bins due to the track momentum resolution. 
The reconstruction efficiency for tracks belonging to a 
true \xe\ bin $i$ is accounted for by factors $\epsilon_{i}^{q}$.
Differences in the slope of the \xe\  spectrum between uds, c and b 
quark events lead to flavour dependent migration effects and to a 
flavour dependent efficiency.
Consequently, the matrix $\xmat_{ij}^{q}$ and $\epsilon_{i}^{q}$ are flavour 
dependent
and have to be applied to the fraction $\fr_{j,k}^q$ of observed
tracks created in a $q = {\rm uds}$, c or b 
event.

These weights $\fr_{j,k}^q$ are the normalised products 
of the flavour dependent fragmentation function $F_{j}^q$ in an
\xe~bin $j$ and the fraction $f_k^q$ of events of a primary quark $q$ in the
considered tag sample:
\begin{equation}
\fr_{j,k}^q = \frac{f_k^qF_{j}^q}
{\sum_{q^\prime} f^{q^\prime}_k F^{q^\prime}_j} .
\label{eq.weighting}
\end{equation}
The  applied weights and the obtained fragmentation functions are strongly
correlated. This was taken into
account in an iterative procedure, whereby the result of the measurement was 
used
to re-calculate $\fr_{j,k}^q$ and to repeat the correction procedure until 
the results were stable. Initial values for the weighting factors
 were taken from Monte Carlo, but alternative initial values were also 
 tried to confirm that the results did not depend on the choice of the
 initial values.
 
The values for $\xmat_{ij}^{q}$ and $\epsilon_{i}^{q}$ were 
obtained from
Monte Carlo. The diagonal elements $\xmat_{ii}^{q}$ of the matrix,
 i.e., the probability that a track measured in its true \xe\ bin is around 
 $80\%$ 
 in most
bins, but becomes significantly smaller for high \xe\ bins in c and b 
flavoured events.
 Values for the efficiency are typically around 
$\epsilon_{i}^{q} \approx 90 \%$ with the exception of the lowest \xe\ bin where 
the efficiency is about 
$50\%$. The efficiency shows only a weak flavour dependence, the values
 differ for different flavours by less than $5\%$.

The corrected number of tracks ${N}^{\rm corrected}_{i,k}$ in each tag sample
 was divided by the corresponding number of tagged hemispheres ${N}^{\rm hemi}_{k}$
 to form a fragmentation function for each tagged sample:
\begin{equation}
{\cal F}_{i,k} = \frac{{N}^{\rm corrected}_{i,k}}{{N}^{\rm hemi}_{k}}.
\end{equation}
\subsection{Flavour tagging and hemisphere correlations}

\Dstar \,mesons with high values of  \xdstar\, and secondary vertices with large
values for the decay length significance \declsig 
are more likely to be found in high energy jets.
 The hemisphere containing the highest energy jet also 
tends to have a higher charged particle multiplicity and a harder track momentum
spectrum than the opposite hemisphere.  Consequently,
 the measured fragmentation functions
in samples with high values for \declsig or \xdstar \ would be too soft.
To correct for this bias, the whole analysis was performed separately for the 
 case
where the tag-hemisphere 
contains the  highest energy jet and where this is not the case. 
The unweighted average of the two results was taken at the end.

The dependence of the track momentum spectrum on the actual value of the decay
length significance and on the \Dstar\ energy was also considered.
Requiring a high value for \declsig or \xdstar \ reduces the phase space for 
gluon
bremsstrahlung and thus introduces a kinematic correlation between 
the hemispheres.
The effect is flavour dependent, becoming more important for higher values of 
\xe\ and is
more pronounced for the \Dstar \,tag than for the secondary vertex tag.
Besides this kinematical effect,
correlations also occur due to geometrical effects: in a typical two jet 
event,
the jets are back to back, thus pointing into geometrically opposite parts of 
the
detector. This introduces a hemisphere correlation if the detector
response is not uniform. In addition to the kinematical and the geometrical 
correlations, the
difference in the fragmentation functions in tagged events and in unselected
events had to be taken into account.

All these effects were accounted for by applying correction factors for 
each tag sample,
flavour and \xe\ bin:
\begin{equation}
 \tagcorr_{i,k}^q
 = \frac{F_{i,k}^q({\rm generated})}
        {F_i^{q}({\rm generated})},
\label{eq.tagcorr}
\end{equation}
i.e., the ratio of the generated fragmentation functions  in tagged events and 
in events where the tag has not been applied.
These correction factors  \tagcorr \ lead to a $10\%$ 
correction for high
\xe \ values  opposite a hemisphere tagged by a secondary vertex and up to a $50\%$
correction for high
\xe \ values  opposite a \Dstar \,tagged hemisphere.
Technically, these factors are applied as corrections to the purities
in the fit procedure, thus taking into account the {\it a priori}
 unknown flavour composition of the tracks in a specific \xe \ bin.  

\subsection{Background subtraction in the \Dstar\ samples}

The measured fragmentation functions in the \Dstar\ signal samples 
${\cal F}_{i,k}^{\rm signal}$ and the side-band samples
${\cal F}_{i,k}^{\rm SB}$ were used to determine
the fragmentation functions for genuine \Dstar\ mesons:
\begin{equation}
 {\cal F}_{i,k}^{\smallDstar}
 = \frac{1}{(1-f^{\rm BG}_k)}
 \left({\cal F}_{i,k}^{\rm signal}
 -f^{\rm BG}_k c_{i,k} {\cal F}_{i,k}^{\rm SB}.
 \right)
\label{eq.corrdstar}
\end{equation}
 The background fractions $f^{\rm BG}_k$ derived from fits to the $\Delta M$
 distribution 
 are listed in table \ref{Table_dstar}. To take into account differences of the
 hemisphere correlations for events in the signal and those in the side-band
 region, correction factors $c_{i,k} = 
 \tagcorr_{i,k}^{\rm signal}/\tagcorr^{\rm SB}_{i,k}$ were applied to the
 fragmentation functions of the side-band samples, i.e., they were multiplied by
 the ratio of the
 correction factors \tagcorr\ for the flavour mix of the signal and the side-band
 sample as predicted by the simulation.

\section{Fits}
\label{section.fits}

A simultaneous fit was performed on the fragmentation functions
${\cal F}_{i,k}^{\smallDstar}$  and ${\cal F}_{i,k}^{\rm vtx}$
of the three \Dstar \,and the three secondary vertex tagged 
samples to extract the flavour dependent fragmentation functions
$F^{\rm uds}$, $F^{\rm c}$ and $F^{\rm b}$.
The  fragmentation functions obtained from samples tagged by  
\Dstar\ decays ${\cal F}_{i,k}^{\smallDstar}$, 
corrected for detector effects and for the combinatorial background
 for each
of the three \xdstar \ regions $k$ and each of the 22 \xe \ bins $i$
were described in the fit by
\begin{equation}
  {\cal F}_{i,k}^{\smallDstar}
  = ({\cal P}^{\rm c}_k \tagcorr_{i,k}^{q={\rm c}})F_{i}^{\rm c}
  + ({\cal P}^{\rm b}_k \tagcorr_{i,k}^{q={\rm b}}) F_{i}^{\rm b},
\label{eqcfit}
\end{equation}
where  the purities ${\cal P}^{\rm c}_k$,
 ${\cal P}^{\rm b}_k$  are given in 
Table
\ref{Table_dstar} and the correction factors $\tagcorr_{i,k}^{q}$
 are defined in
equation \ref{eq.tagcorr}.
Na\"{\i}vely, the fragmentation function corrected for detector effects
 for each
of the three samples tagged by secondary vertices 
and each of the 22 \xe \ bins could be
 described by
\begin{equation}
{\cal F}_{i,k}^{\rm vtx}
=   (f_{k}^{uds} \tagcorr_{i,k}^{q={\rm uds}} )F_i^{\rm uds} 
  + (f_{k}^{c} \tagcorr_{i,k}^{q={\rm c}}) F_i^{\rm c} 
  + (f_{k}^{b} \tagcorr_{i,k}^{q={\rm b}}) F_i^{\rm b}.
                        \label{eq_naive_bfit}
\end{equation}
However, the fraction of c events $f_{k}^{\rm c}$ in all three vertex tagged samples
is small (see Table \ref{Table1}), hence
 the relative uncertainty on these fractions large. If
 equation \ref{eq_naive_bfit} would be used,
 the vertex tagged samples would dominate the fit results  
 due to their larger statistical weight as compared to the
 \Dstar\ tagged samples. To ensure that the
  \Dstar\ samples are used to obtain the charm fragmentation function, the 
  flavour fraction $f_k^{\rm uds}$ and $f_{k}^{\rm c}$ were replaced by 
  $(1- f_k^{b})R_{\rm uds}/(R_{\rm uds}+R_{\rm c})$ and
  $(1- f_k^{b})R_{\rm c}/(R_{\rm uds}+R_{\rm c})$ where $R_{\rm uds}$ and
  $R_{\rm c}$ 
  are the Standard Model values for the branching
fractions $R_q = \Gamma(\PZz\rightarrow \Pq\Paq)/
\Gamma(\PZz\rightarrow {\rm hadrons})$. The decay length dependence
of the ratio of uds to c events were accounted for by correction factors
$d_{i,k}$ to the measured fragmentation function.
 The fragmentation function of the secondary vertex tagged 
 samples was then described by
\begin{equation}
d_{i,k} {\cal F}_{i,k}^{\rm vtx}
= (1-(f_k^{\rm b} \tagcorr_{i,k}^{q={\rm b}} ))F_i^{\rm udsc} +
                        (f_{k}^{\rm b} \tagcorr_{i,k}^{q={\rm b}}) F_i^{\rm b},
                        \label{eqbfit}
\end{equation}
where 
\begin{equation}
F^{\rm udsc} = 
 \frac{  R_{\rm uds}F_i^{\rm uds} 
      + R_{\rm c}  F_i^{\rm c}   }
      { R_{\rm uds} + R_{\rm c} }.
                        \label{equdsfit}
\end{equation}
The correction factors $d_{i,k}$ were 
derived from the momentum spectrum in Monte Carlo events with the
ratio of uds to c events taken to be the same in all vertex samples, 
divided by the unmodified 
 momentum spectrum with a variable ratio of uds to c events. 
 These correction are of the order of $1\%$ for
most of the \xe\ bins except for the highest \xe\ bins 
where they exceed $10\%$.

\begin{figure}[p]
\begin{center}
\resizebox{160mm}{!}{\includegraphics*{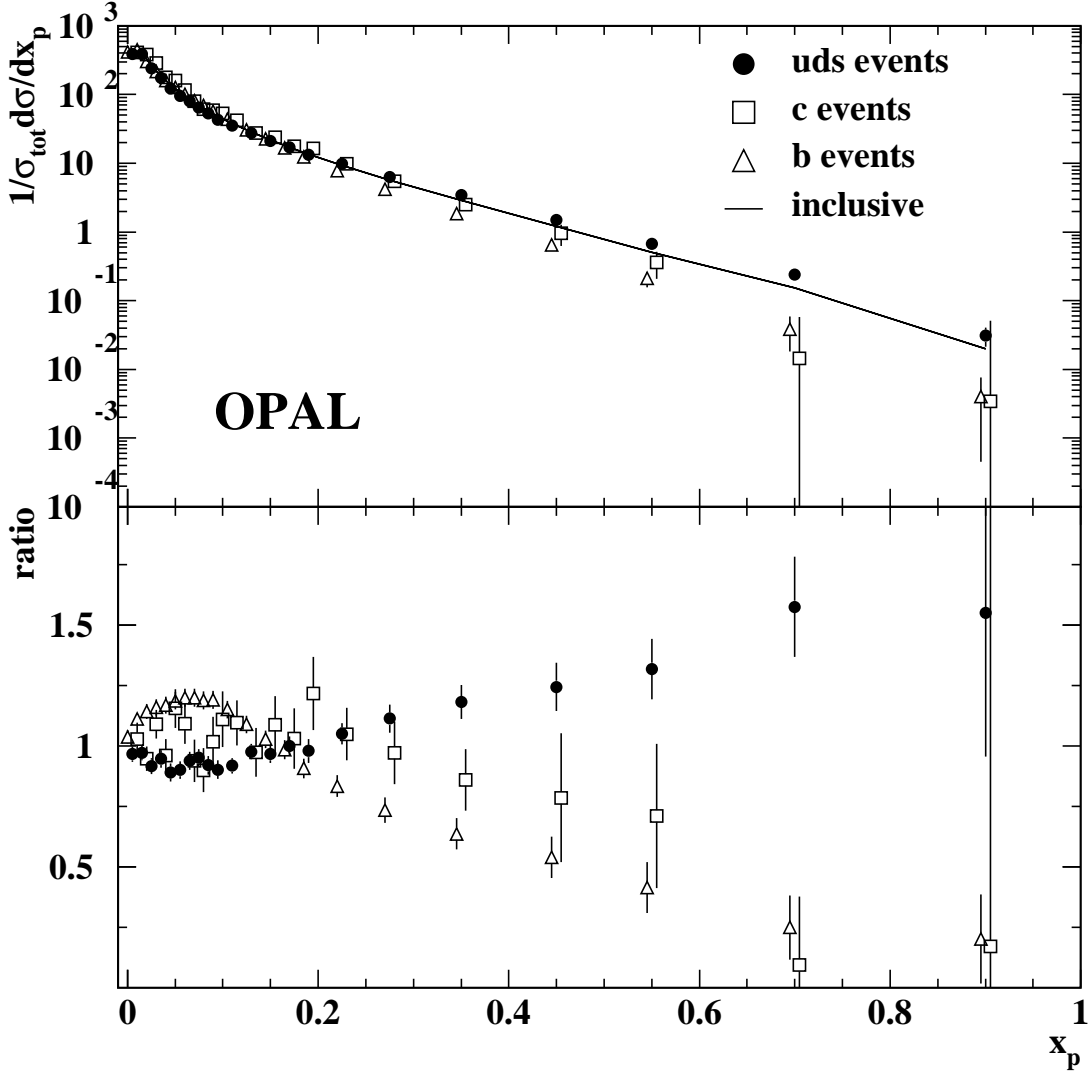}}
\end{center}
\caption[ ]{\sl The upper plot shows the
measured fragmentation functions for uds events (filled symbols), c events
(open squares) and b events (open triangles) as well as the inclusive
 fragmentation function (solid line). The lower plot shows 
the ratio of the flavour dependent fragmentation functions to the inclusive 
fragmentation 
function. The error bars include statistical and systematic uncertainties.
The systematic uncertainties are correlated between bins as well as between
flavours. 
\label{plot4}}
\end{figure}

A simultaneous fit was performed to
extract $F^{\rm uds}$, $F^{\rm c}$ and
$F^{\rm b}$. In fact, the secondary vertex data using equation 
\ref{eqbfit} essentially fixes $F^{\rm udsc}$ and 
$F^{\rm b}$ and then the \Dstar \,data provide
$F^{\rm c}$ through equation \ref{eqcfit}, allowing equation
\ref{equdsfit} to give $F^{\rm uds}$.
The fit was based on the track momentum spectrum of the hemisphere opposite
the tag.
Therefore, the results had to be multiplied by a factor of two to obtain
the full event fragmentation functions as they are shown in Fig.\,\ref{plot4} and in 
Table \ref{Table_uds}. The mean values of these distributions
and their statistical uncertainty are:

\begin{center}
\begin{tabular}{ l @{ } c @{ } c @{ } c@{ } l@{ }}
$\xemean^{\rm uds}$ & $ = $&$ 0.0630  $&$\pm$&$ 0.0003 $ \\
$\xemean^{\rm c}$   & $ = $&$ 0.0576  $&$\pm$&$ 0.0012 $ \\
$\xemean^{\rm b}$   & $ = $&$ 0.0529  $&$\pm$&$ 0.0001. $ \\
\end{tabular}
\end{center}
 
The results can be compared with results for the inclusive fragmentation function
which was obtained from the track momentum spectrum of all events without
considering any flavour tagging. These results are shown in Table
\ref{Table_tot} and the mean value of the distribution was found to be:
\begin{center}
\begin{tabular}{ l @{ } c @{ } c @{ } c@{ } l}
$ \xemean^{\rm incl}$ & $ 
 = $&$ 0.05938  $&$\pm$&$ 0.00002 $.  \\
\end{tabular}
\end{center}

The results for the \xip\ distribution are shown if Fig.\,\ref{plot9}
and Table \ref{Table_xi_uds}. To determine the positions of the maxima, \ximax,
skewed Gaussians, i.e., combinations of two Gaussians with different widths
to the left and to the right from the centre were fitted to these distribution 
as motivated by the next-to-leading-log (NLLA) approximation
\cite{xi_nlla}. Following the procedure in
 \cite{opal133}, the fit was performed in the
region $2.2<\xip<5.0$. The results for the positions of the maxima with their
statistical uncertainties are:

\begin{center}
\begin{tabular}{ l @{ } c @{ } c@{ } c@{ } l@{ } }
$ \ximax^{\rm uds}$ & $  = $&$ 3.74  $&$\pm$&$ 0.06 $\\
$ \ximax^{\rm c}   $ & $  = $&$ 3.63  $&$\pm$&$ 0.16 $ \\
$ \ximax^{\rm b}   $ & $  = $&$ 3.55  $&$\pm$&$ 0.01 $.  \\
\end{tabular}
\end{center}

Again, these flavour dependent results can be compared with the results of the 
inclusive \xip\ distribution as obtained from all events without any flavour
tagging. The results are shown in Table \ref{Table_xi_tot}, the position of
the maximum was determined to be:
\begin{center}
\begin{tabular}{ l @{ } c @{ } c@{ } c@{ } l@{ } }
$ \ximax^{\rm incl} $ & $  = $&$ 3.656 $&$\pm$&$ 0.003$ .\\
\end{tabular}
\end{center}
 
\section{Systematic errors and cross checks}
%
The systematic uncertainties affecting the above results are due to the following
sources: (1) uncertainties of  the purities of the samples tagged by secondary
vertices and (2) by   \Dstar\,decays,  the 
\Dstar \,reconstruction (3), the 
hemisphere correlation (4),  uncertainties inherent to the correction
procedure (5) and the
track and event selection (6). These were estimated from the difference 
between the
central value and the result of the repeated analysis after a
 cut, a purity or the correction procedure was modified. In each case, the largest
 deviation was taken as the systematic error. 

\begin{itemize}

\item[(1)]
The uncertainties on the purities in the secondary vertex event samples
were estimated using the published results from the measurement of 
the charged multiplicity in b, c and uds events \cite{opalc}.
There, variations of the measured multiplicity due to the uncertainties in the
b lifetime, the fragmentation of b and c quark events, the production rates and
 the
mixture of b hadrons produced as well as the decay multiplicities
were studied and can be used to derive the uncertainties on the purities.
The secondary vertex sample purities were then varied in the range of their
 uncertainty and
the resulting differences of the results for the fragmentation function 
was taken as the systematic error due to this source.

\item[(2)]
The purities ${\cal P}^{\rm c}_k$ and ${\cal P}^{\rm b}_k$ of the \Dstar \, 
tag bins were taken from \cite{opaldstar}. They
were modified
within their systematic errors  to obtain the contribution 
to the systematic uncertainty on the
fragmentation function.

\item[(3)]
To study the impact of the details of the \Dstar 
\,candidate
selection, the analysis was repeated with four sets of modified selection
criteria.
 The $M_{\smallD0}^{\rm cand}$ mass window
 was increased to
$1.765\ \gevcc < M_{\smallD0}^{\rm cand} < 1.965\ \gevcc$; the $\Delta M$ window 
was
increased to $0.141\ \gevcc< \Delta M < 0.150\ \gevcc$; only one of the three 
tracks
was required to have a $z$-chamber or jet chamber end point $z$ measurement and
the cuts based on ${\rm d}E/{\rm d}x$ were removed. The last modification
lead to a reduction of the signal to
noise ratio of more than $25 \%$.

\item[(4)]
The correlation between hemispheres 
caused by the kinematic of gluon radiation is corrected for by the factors 
\tagcorr (equation \ref{eq.tagcorr}) .
A good description of the energy spectrum and the angular distribution of jets
in the Monte Carlo simulation is important for a reliable prediction of this effect.
To estimate the effects of small discrepancies between data and simulation
in these distribution, the analysis was repeated  applying weights to 
Monte Carlo events so that the 
energy distribution of the
most energetic jet and the distribution of the angle between the two most
energetic jets in data and Monte Carlo simulation agreed. Most of the resulting event 
weights had values
between 0.95 and 1.05.
The difference of the results with and without weighting 
was taken as the systematic uncertainty.

\item[(5)]
To estimate systematic uncertainties due to the correction for track momentum
resolution and efficiency (Section \ref{corr.track}) and corrections applied in the
fitting procedure (Section \ref{section.fits}), the following two modifications to
the correction procedure were tested.
First, the weighting factors $\fr_{j,k}^q$ as defined in
equation \ref{eq.weighting} were not re-calculated in an iterative procedure
but were based on the initial values from the Monte Carlo simulation.
Secondly, the correction factors $d_{i,k}$  in equation \ref{eqbfit}
were omitted, i.e., the  secondary vertex samples were not corrected for the 
 variation of the uds to c flavour fraction.

\item[(6)]
To account for imperfections in the tracking detector simulation, 
results were obtained in six different ways with modified
event and track selections and variations of track quantities in the
Monte Carlo simulation.
The cut on the angle of the thrust axis 
$|\cos \theta_{\rm thrust}| > 0.8$ was removed; instead of
accepting all tracks, a cut on $|\cos \theta_{\rm track}| < 0.7$  was applied;
tracks were rejected if their $z$-coordinate at the point of closest approach
to the event origin was larger than $|z_0| > 10$\,cm; tracks were rejected
if their momentum was smaller than 0.250 \gevc; the track momenta in simulated
events were modified by an additional smearing factor leading to a degradation of
the momentum resolution in Monte Carlo events of $10\%$; 
the simulated track momenta were shifted by $1\%$.
Since these effects are flavour independent, 
the uncertainty due to the track and event selection has been set to
be the same for the inclusive and the flavour dependent  
fragmentation functions.

\end{itemize}

The  systematic uncertainties from the above groups of effects were
added in quadrature and are shown for the flavour dependent distributions
in the last column of Tables \ref{Table_uds} and \ref{Table_xi_uds}.
The result for the inclusive fragmentation function was obtained without flavour 
tagging and
consequently does not depend on the tagging efficiency, purity or hemisphere 
correlation, so only the last two groups of effects contributed to the systematic
uncertainty shown in the last column of Tables \ref{Table_tot} and 
\ref{Table_xi_tot}.  
For most of the \xe\ range, the relative systematic error
is below $5 \%$ for uds and b events and below $10 \%$ for c events. For
very high momenta ($\xe > 0.5$), the systematic uncertainty
becomes larger. Note that in uds and c events, the systematic 
and statistical errors are roughly equal.
 Detailed results are shown in Table \ref{tabsys1}  for 
a typical low momentum \xe\ bin ($0.05 < \xe < 0.06$)
and a typical high momentum \xe\ bin ($0.3 < \xe < 0.4$).

In Table \ref{tabsys2}, the systematic uncertainties
 on the measurement of the mean value \xemean\ of
the \xe\ distributions are shown. Further systematic 
checks were done for the determination of the position
of the maximum of the \xip\ distributions, \ximax. Instead of evaluating the
 position of the maximum using a skewed Gaussian fit, a normal Gaussian fit
 as motivated by LLA \cite{xi_lla} was used. Furthermore, the
fit range was modified and the skewed Gaussian was fitted to the
 measured \xip\
distribution in the regions $2.0<\xip<5.2$ and $2.4<\xip<4.8$. 
 The uncertainty obtained from this test 
 has been added in quadrature to the previous six 
contributions as listed in Table \ref{tabsys3}.

To cross-check the results, the tagging  methods were modified. An
alternative b tag was applied, based on impact parameter information
rather than on decay length information: The third largest impact
parameter of a track was taken as the tag quantity. The impact parameter
distribution of tracks from a decay are independent from the
energy of the decaying particle while the decay length of a particle
is proportional to its energy. Hence, this simple alternative method is
less  affected by kinematical correlations due to gluon radiation but 
affected by other systematic effects than the standard method. The
result obtained with the alternative b tagging method is consistent with the
central values 
within the assigned systematic error. Another cross-check was performed using an 
alternative background 
treatment in the
\Dstar \,tagged event samples. 
Instead of taking the \Dstar \,purities from
\cite{opaldstar} and subtracting the effect of the combinatorial
 background with the help
of a fit to the \Dstar \,signal, purities and background were taken from
Monte Carlo. Again, the results obtained with this alternative
method and the central values agreed within the 
systematic errors.

Effects from the binning in the tag variable have also been cross-checked.
Instead of using three secondary vertex bins and three
 \Dstar\
bins, alternative results were obtained using two and four secondary 
vertex bins and likewise two and four \Dstar\ bins. The deviations
from the central value were in all cases smaller than the estimated 
systematic  error.

The whole analysis procedure has also been tested 
globally with simulated events. It was shown
that the generated fragmentation function and the result of 
the unfolding procedure agree within the statistical uncertainty.

By integrating the fragmentation functions, the charged multiplicity
in uds, c and b events can be obtained. The results,
\begin{center}
\begin{tabular}{l @{ } c @{ } c@{ } c@{ } c@{ } c@{ }c@{\,}c}
$ n^{\rm uds}$ & $  = $&$  20.25 $&$\pm$&$  0.11 $&$\pm$&$0.37$&\\
$ n^{\rm c}  $ & $  = $&$  21.55 $&$\pm$&$  0.37 $&$\pm$&$0.64$&\\
$ n^{\rm b}  $ & $  = $&$  23.16 $&$\pm$&$  0.02 $&$\pm$&$0.45$&\\
$ n^{\rm incl}$ & $  = $&$  21.16 $&$\pm$&$  0.01 $&$\pm$&$0.21$&,\\
\end{tabular}
\end{center}
are in good agreement with results of direct measurements of the
charged multiplicities \cite{opalb,opalc,opallff,sld}.

The  average of the 
three flavour dependent fragmentation functions can be formed,
weighted with the Standard Model branching fractions 
$R_{\rm uds},R_{\rm c}$
and $R_{\rm b}$. This combined fragmentation function can be 
compared with the  results for the inclusive
fragmentation function and with previously published OPAL results 
\cite{opallff}. All three results show good agreement with each other.

\section{Results}

The results for the flavour dependent fragmentation functions 
for uds, c and b events as well as the  inclusive fragmentation function
are shown in Fig.\,\ref{plot4} and in Tables
\ref{Table_uds} and \ref{Table_tot}. 
The mean values of these distributions are: 

\begin{center}
\begin{tabular}{l @{ } c @{ } c @{ } c@{ } c@{ } c@{ }c@{\,}c}
$ \xemean^{\rm uds}$ & $ 
 = $&$ 0.0630  $&$\pm$&$ 0.0003 $&$\pm$&$ 0.0011 $& \\
$ \xemean^{\rm c}$   & $ 
 = $&$ 0.0576  $&$\pm$&$ 0.0012 $&$\pm$&$ 0.0016 $& \\
$ \xemean^{\rm b}$   & $ 
 = $&$ 0.0529  $&$\pm$&$ 0.0001 $&$\pm$&$ 0.0013 $&  \\
$ \xemean^{\rm incl}$ & $ 
 = $&$ 0.05938  $&$\pm$&$ 0.00002 $&$\pm$&$ 0.00057$&.  \\
\end{tabular}
\end{center} 

The light quark fragmentation function is found to be harder than the 
b quark fragmentation function as expected due to the cascade decays 
of b hadrons in b quark events with more particles sharing the energy. 
This observation is also consistent with the results of comparisons
of gluon, uds and b jets \cite{gluonb1,gluonb2}.

\begin{figure}[p]
\begin{center}
\resizebox{160mm}{!}{\includegraphics*{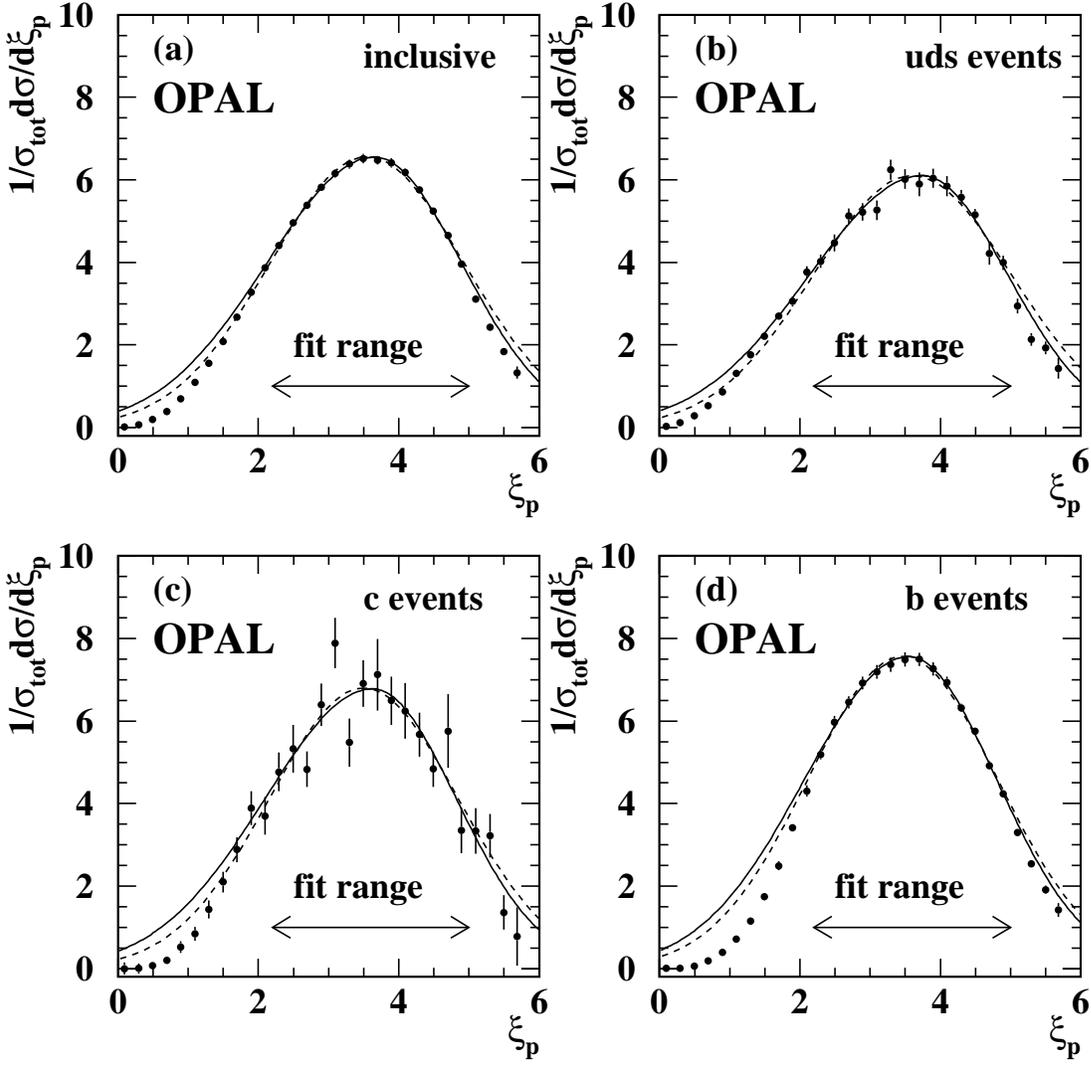}}
\end{center}
\caption[ ]{\sl 
$\xip = \ln(1/\xe)$ distribution for (a) all events, (b) uds events, (c) c events and (d) b 
events.
The solid lines show the results of the
skewed Gaussian fitted to the distributions in the indicated fit range and
the dashed lines show the results of a normal Gaussian fit.
The error bars include statistical and systematic uncertainties.
\label{plot9}}
\end{figure}

In Fig.\,\ref{plot9} and in Tables \ref{Table_xi_uds} and \ref{Table_xi_tot},
the results are presented for the $\xip = \ln(1/\xe)$ distribution
which emphasises the lower momenta of the spectrum. 
Skewed Gaussians were fitted to these distributions to 
obtain the position of their maxima:
\begin{center} 
\begin{tabular}{l @{ } c @{ } c@{ } c@{ } c@{ } c@{ }c@{\,}c}
$ \ximax^{\rm uds}$ & $  = $&$ 3.74  $&$\pm$&$ 0.06 $&$\pm$&$ 0.21  $&\\
$ \ximax^{\rm c}   $ & $  = $&$ 3.63  $&$\pm$&$ 0.16 $&$\pm$&$ 0.31 $&\\
$ \ximax^{\rm b}   $ & $  = $&$ 3.55  $&$\pm$&$ 0.01 $&$\pm$&$ 0.07 $&  \\
$ \ximax^{\rm incl} $ & $  = $&$ 3.656 $&$\pm$&$ 0.003$&$\pm$&$ 0.115$&.  \\
\end{tabular}
\end{center} 
The result for the inclusive distribution is in good agreement with 
previous
results \cite{xiresults,sameE1}, whereas the position of the maxima 
of the  flavour dependent \xip\ distribution is reported here 
 for the first time.
Part of the systematic uncertainties cancel when the ratio of the flavour
 dependent
results to the inclusive result is taken:
\begin{center} 
\begin{tabular}{ c @{ } c @{ } c@{ } c@{ } c@{ } c@{ }c@{\,}c}
$ \ximax^{\rm uds}/\ximax^{\rm incl}$ & $  = $&
          $ 1.023  $&$\pm$&$ 0.017 $&$\pm$&$0.028 $& \\
$ \ximax^{\rm c}/\ximax^{\rm incl}   $ & $  = $&
          $ 0.993  $&$\pm$&$ 0.044 $&$\pm$&$ 0.082 $& \\
$ \ximax^{\rm b}/\ximax^{\rm incl}   $ & $  = $&
          $ 0.971 $&$\pm$&$ 0.003 $&$\pm$&$ 0.022 $&. \\
\end{tabular}
\end{center}
Another indication for a flavour dependence of the
\xip\ distribution is given by the differences of the shape of the
distributions in figure \ref{plot9}.

\begin{figure}[p]
\begin{center}
\resizebox{160mm}{!}{\includegraphics*{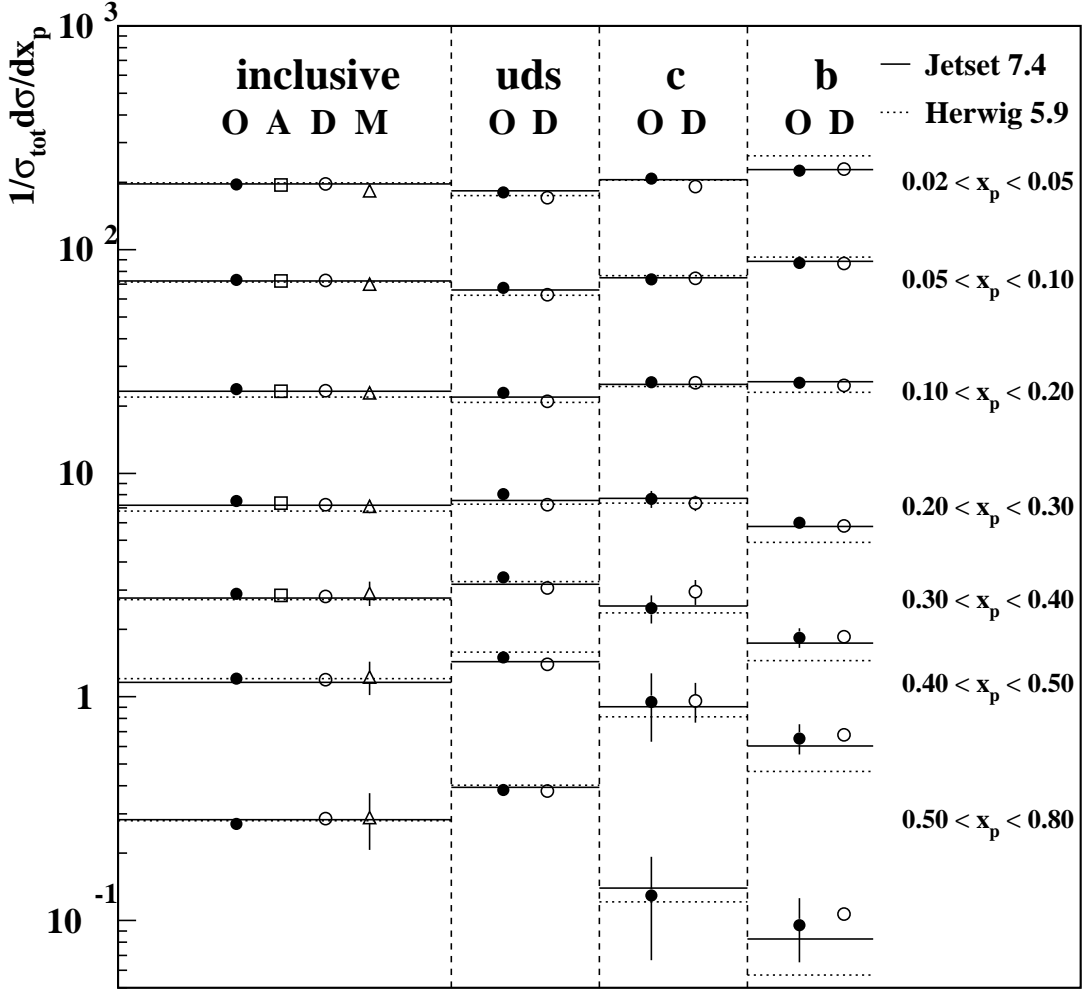}}
\end{center}
\caption[ ]{\sl 
Comparison of the results for the inclusive fragmentation function for this analysis (O)
with results from ALEPH (A), DELPHI (D) and MARK~II (M) at 
$\sqrt{s}=m_{{\rm Z}^0}$  
 \cite{sameE1,sameE2} and of the flavour dependent fragmentation function with the results 
 from DELPHI \cite{delphi2}.
The error bars include statistical and systematic uncertainties.
 The Jetset 7.4 predictions for the fragmentation function are shown 
 as full horizontal lines and the Herwig 5.9 predictions as dotted 
 horizontal lines.
\label{plot7}}
\end{figure}

In Fig.\,\ref{plot7}, the  total and the flavour dependent 
fragmentation functions
are compared with results from other experiments at the same centre-of-mass 
energy. There is good agreement with  \cite{delphi2} where the
 $\Dstarp \rightarrow {\rm K}^-\pi^+\pi^+$ decay was used as well, 
but a different b tagging method and a different
correction procedure was applied. Also the quoted systematic uncertainty is similar
in size with the exception of the high momentum region, where the uncertainty in 
\cite{delphi2} is smaller than in this paper. However, a direct comparison of
the systematic errors is difficult since the
error sources dominating the systematic uncertainty 
in the high momentum region in this paper (Table \ref{tabsys1})
are not explicitly considered in 
\cite{delphi2}.
Also in Fig.\,\ref{plot7}, the results are shown to be consistent 
with the Jetset 7.4 expectation
while the Herwig 5.9 Monte Carlo program
\cite{herwig}\footnote{The parameter set used 
was the same as in \cite{gluonb1} for Herwig 5.8, except for the value of 
the cluster mass
cutoff CLMAX which has been increased from 3.40 \gevcc\ to 3.75 \gevcc. 
Alternatively, studies were
done with the default parameter set of the Herwig Monte Carlo program. But in 
this case, even the inclusive fragmentation function
fails to describe the data.} 
fails to describe fully the b fragmentation function.

\begin{figure}[p]
\begin{center}
\resizebox{160mm}{!}{\includegraphics*{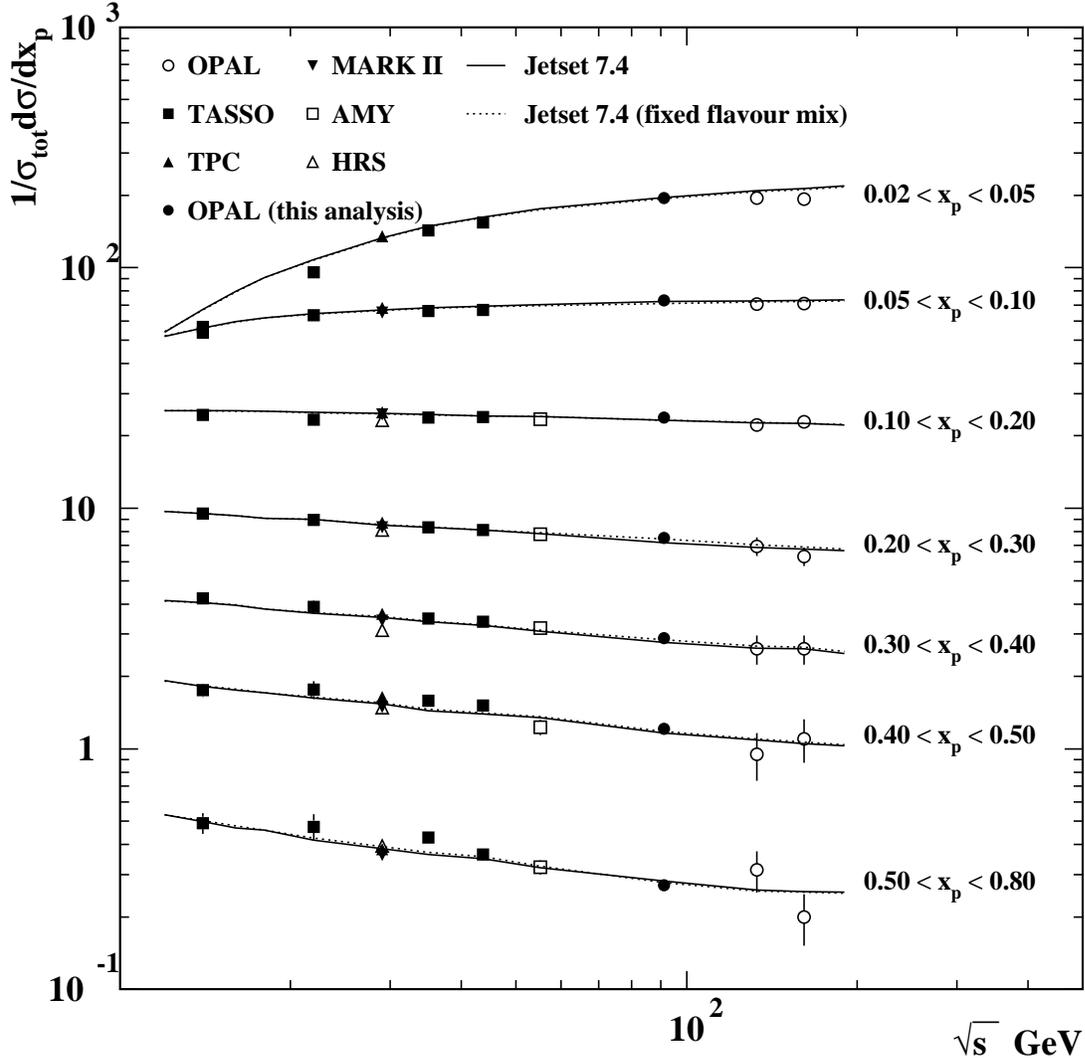}}
\end{center}
\caption[ ]{\sl 
Comparison of the results for the inclusive fragmentation function 
with results at different lower \cite{lowE}
and higher  centre-of-mass energies \cite{highE,opal133}. 
The error bars include statistical and systematic uncertainties.
The solid lines show
the Jetset 7.4 prediction, assuming the 
centre-of-mass energy dependence of the flavour composition as predicted by the
electroweak theory. The dotted lines show the Jetset 7.4 prediction assuming
 for all
energies the same flavour mix as at $\sqrt{s} = 18$ \gev. The dotted line
is almost entirely hidden behind the full line and even at $\sqrt{s} =
91.2 $ \gev, only a negligible difference between the two curves can be seen
because the effect of an increased b contribution is compensated largely 
by the effect of a decreased c contribution.
\label{plot6}}
\end{figure}

The measurement of the total fragmentation
function in comparison to measurements at lower energies \cite{lowE} 
and at centre-of-mass
energies between $130~\gev$ and $161~\gev$
 \cite{highE,opal133} 
is shown in Fig.\,\ref{plot6}. Apart from the lowest \xe\ region,
there is good agreement with the Jetset 7.4 prediction 
(solid line) despite the fact that the parameters used in Jetset 7.4
 were optimised to describe data
at the \z\ resonance.
In the highest \xe\
bin, the difference between the  
total fragmentation function  measured at $\sqrt{s}=14.0~\gev$ and at $\sqrt{s}=161~\gev$
is of the same order of magnitude as 
the difference between uds and b fragmentation function at 
$\sqrt{s}=91.2~\gev$. Observing the good agreement between data and the Jetset 7.4
 prediction of the
$\sqrt{s}$ dependence of the total fragmentation function and of the flavour dependent
fragmentation functions at $\sqrt{s}=91.2~\gev$,
 we can use Jetset to estimate the effect of the
change of the flavour mix to the apparent scaling violation.
In Fig.\,\ref{plot6}, the Jetset 7.4 prediction is shown when
 the flavour mix at all centre-of-mass energies was forced to be the same as
the flavour mix at $\sqrt{s}=18~\gev$. Although this 
fixed flavour mix
is very different from that at at the \z\ peak
and the flavour dependent fragmentation
functions differ significantly, the changes on the total fragmentation function 
at $\sqrt{s}=91.2~\gev$ are of the order of only two to four percent.
 This is due to a
cancellation of the effect of an increased b contribution and a decreased c 
contribution
at centre-of-mass energies close to the \z\ resonance.

Comparing the positions of
the maxima, \ximax, of the \xip\ distribution, the flavour dependence is less 
pronounced 
 than for the fragmentation function at 
high \xe. The values for \ximax\ at 
$\sqrt{s}=14.0~\gev$ 
and $\sqrt{s}=161~\gev$ differ by almost a factor of two, while the 
 difference between 
the flavour dependent results at $\sqrt{s}=91.2~\gev$ 
is one order of magnitude smaller. 
Like in the case of the fragmentation functions,  an increased b 
contribution
 and a simultaneously decreased c contribution leads to cancellations 
 when comparing low energy measurements and results at the \z\ peak.
This is expected on the basis of simulations where the \ximax\ obtained  for the
flavour mixture at $\sqrt{s}=91.2~\gev$ and at 
$\sqrt{s}=18.0~\gev$ differ by less than one percent.

\section{Conclusions}

Flavour dependent fragmentation functions in $\z \rightarrow 
{\rm q}\bar {\rm q}$
events have been measured separately for bottom, charm and light (uds) quarks 
and as well as for all flavours together.
These measurements are based on OPAL data recorded between 1990 and 1995.
Event samples with different flavour compositions were formed using
reconstructed \Dstar \,mesons and
secondary vertices in jets. The charged particle momentum spectrum has been studied
in the event hemisphere opposite to the tag. A simultaneous fit was performed to
extract the flavour dependent \xe\ distribution 
as well as the flavour dependent \xip\
distribution.

 The fragmentation function for b quarks is significantly
softer than for uds quarks. The fragmentation functions are well described
by the Jetset 7.4 Monte Carlo program while Herwig 5.9 fails to describe fully the
b fragmentation function.

For the first time, flavour dependent \xip\ distributions have been studied. The
flavour dependence of the position of the maximum has been determined and was found
to be small compared with the differences of this value 
at different centre-of-mass energies. 

\section*{Acknowledgements}

We particularly wish to thank the SL Division for the efficient operation
of the LEP accelerator at all energies
 and for their continuing close cooperation with
our experimental group.  We thank our colleagues from CEA, DAPNIA/SPP,
CE-Saclay for their efforts over the years on the time-of-flight and trigger
systems which we continue to use.  In addition to the support staff at our own
institutions we are pleased to acknowledge the  \\
Department of Energy, USA, \\
National Science Foundation, USA, \\
Particle Physics and Astronomy Research Council, UK, \\
Natural Sciences and Engineering Research Council, Canada, \\
Israel Science Foundation, administered by the Israel
Academy of Science and Humanities, \\
Minerva Gesellschaft, \\
Benoziyo Center for High Energy Physics,\\
Japanese Ministry of Education, Science and Culture (the
Monbusho) and a grant under the Monbusho International
Science Research Program,\\
German Israeli Bi-national Science Foundation (GIF), \\
Bundesministerium f\"ur Bildung, Wissenschaft,
Forschung und Technologie, Germany, \\
National Research Council of Canada, \\
Research Corporation, USA,\\
Hungarian Foundation for Scientific Research, OTKA T-016660, 
T023793 and OTKA F-023259.\\

\newpage

%
%
%

%
%
\begin{table}[p]
\begin{center}
\begin{tabular}{|c|c@{}c@{}c@{}c@{}c|c@{} c@{} c@{} c@{} c|c@{}  c @{}c@{} c@{} c|}   
                            \hline
   \xe   &  \multicolumn{15}{|c|}
       {$1/{\sigma_{\mathrm{tot}}}\cdot{\deriv{\sigma^h}}/{\deriv{\xE}}$}  \\
       \cline{2-16}
   & \multicolumn{5}{|c|}{uds events}  
   & \multicolumn{5}{|c|}{c events} 
   & \multicolumn{5}{|c|}{b events}\\
                            \hline
                            \hline
%
%
0.00--0.01
&388.&$\pm$&  5.&$\pm$&  9.
&413.&$\pm$& 19.&$\pm$& 18.
&416.&$\pm$&  1.&$\pm$&  8.
\\
0.01--0.02
&390.&$\pm$&  5.&$\pm$& 10.
&381.&$\pm$& 17.&$\pm$& 11.
&447.&$\pm$&  1.&$\pm$&  8.
\\
0.02--0.03
&241.&$\pm$&  4.&$\pm$&  7.
&287.&$\pm$& 13.&$\pm$&  8.
&300.&$\pm$&  1.&$\pm$&  7.
\\
0.03--0.04
&176.&$\pm$&  3.&$\pm$&  5.
&178.&$\pm$& 11.&$\pm$&  6.
&215.&$\pm$&  1.&$\pm$&  5.
\\
0.04--0.05
&122.6&$\pm$&  2.7&$\pm$&  3.9
&159.&$\pm$&  10.&$\pm$&  5.
&160.7&$\pm$&  0.6&$\pm$&  4.1
\\
0.05--0.06
& 95.7&$\pm$&  2.2&$\pm$&  2.9
&116.&$\pm$&  8.&$\pm$&  4.
&126.1&$\pm$&  0.5&$\pm$&  3.4
\\
0.06--0.07
& 79.3&$\pm$&  1.9&$\pm$&  2.3
& 79.&$\pm$&  7.&$\pm$&  3.
&101.4&$\pm$&  0.4&$\pm$&  2.7
\\
0.07--0.08
& 65.0&$\pm$&  1.6&$\pm$&  1.7
& 61.&$\pm$&  6.&$\pm$&  2.
& 81.9&$\pm$&  0.4&$\pm$&  2.2
\\
0.08--0.09
& 53.3&$\pm$&  1.6&$\pm$&  1.3
& 59.&$\pm$&  6.&$\pm$&  2.
& 68.9&$\pm$&  0.4&$\pm$&  1.9
\\
0.09--0.10
& 43.3&$\pm$&  1.5&$\pm$&  1.0
& 53.&$\pm$&  5.&$\pm$&  2.
& 57.1&$\pm$&  0.3&$\pm$&  1.6
\\
0.10--0.12
& 35.1&$\pm$&  0.9&$\pm$&  0.7
& 41.9&$\pm$&  3.2&$\pm$&  1.5
& 44.0&$\pm$&  0.2&$\pm$&  1.3
\\
0.12--0.14
& 27.7&$\pm$&  0.7&$\pm$&  0.4
& 27.6&$\pm$&  2.6&$\pm$&  1.2
& 30.9&$\pm$&  0.2&$\pm$&  1.0
\\
0.14--0.16
& 21.2&$\pm$&  0.7&$\pm$&  0.4
& 23.8 &$\pm$&   2.4&$\pm$&  1.0
& 22.5&$\pm$&  0.1&$\pm$&  0.8
\\
0.16--0.18
& 17.1&$\pm$&  0.6&$\pm$&  0.3
& 17.6&$\pm$&  2.0&$\pm$&  0.8
& 16.8&$\pm$&  0.1&$\pm$&  0.6
\\
0.18--0.20
& 13.3&$\pm$&  0.6&$\pm$&  0.3
& 16.5&$\pm$&  1.9&$\pm$&  0.7
& 12.3&$\pm$&  0.1&$\pm$&  0.5
\\
0.20--0.25
&  9.86&$\pm$&  0.26&$\pm$&  0.30
&  9.8&$\pm$&  0.9&$\pm$&  0.5
&  7.82&$\pm$&  0.05&$\pm$&  0.40
\\
0.25--0.30
&  6.30&$\pm$&  0.19&$\pm$&  0.25
&  5.5&$\pm$&  0.7&$\pm$&  0.3
&  4.16&$\pm$&  0.04&$\pm$&  0.29
\\
0.30--0.40
&  3.42&$\pm$&  0.09&$\pm$&  0.17
&  2.49&$\pm$&  0.31&$\pm$&  0.19
&  1.84&$\pm$&  0.02&$\pm$&  0.18
\\
0.40--0.50
&  1.50&$\pm$&  0.05&$\pm$&  0.10
&  0.95&$\pm$&  0.19&$\pm$&  0.26
&  0.65&$\pm$&  0.01&$\pm$&  0.10
\\
0.50--0.60
&  0.668&$\pm$&  0.033&$\pm$&  0.048
&  0.36&$\pm$&  0.11&$\pm$&  0.10
&  0.210&$\pm$&  0.006&$\pm$&  0.052
\\
0.60--0.80
&  0.241&$\pm$&  0.008&$\pm$&  0.024
&  0.014&$\pm$&  0.041&$\pm$&  0.015
&  0.038&$\pm$&  0.001&$\pm$&  0.020
\\
0.80--1.00
&  0.031&$\pm$&  0.007&$\pm$&  0.007
&  0.003&$\pm$&  0.046&$\pm$&  0.012
&  0.0040&$\pm$&  0.0005&$\pm$&  0.0035
\\
%
%
                                           \hline
\end{tabular}
\caption{\sl Fragmentation functions of uds, c and b events. The first error is statistical, the second 
systematic.
\label{Table_uds}}
\end{center}
\end{table} 
%
%
\begin{table}[p]
\begin{center}
\begin{tabular}{|c|c@{ }c@{ }c@{ }c@{ }c|}   
                            \hline
   \xe   &  \multicolumn{5}{|c|}
       {$1/{\sigma_{\mathrm{tot}}}\cdot{\deriv{\sigma^h}}/{\deriv{\xE}}$}  \\
                            \hline
                            \hline
%
%
0.00--0.01&401.2&$\pm$&  0.3&$\pm$&  7.4\\
0.01--0.02&401.6&$\pm$&  0.2&$\pm$&  4.9\\
0.02--0.03&262.8&$\pm$&  0.2&$\pm$&  3.9\\
0.03--0.04&185.5&$\pm$&  0.2&$\pm$&  2.9\\
0.04--0.05&137.5&$\pm$&  0.1&$\pm$&  2.1\\
0.05--0.06&106.2&$\pm$&  0.1&$\pm$&  1.6\\
0.06--0.07& 84.5&$\pm$&  0.1&$\pm$&  1.2\\
0.07--0.08& 68.3&$\pm$&  0.1&$\pm$&  0.9\\
0.08--0.09& 57.9&$\pm$&  0.1&$\pm$&  0.7\\
0.09--0.10& 47.9&$\pm$&  0.1&$\pm$&  0.6\\
0.10--0.12& 38.19&$\pm$&  0.05&$\pm$&  0.42\\
0.12--0.14& 28.30&$\pm$&  0.04&$\pm$&  0.28\\
0.14--0.16& 21.88&$\pm$&  0.03&$\pm$&  0.19\\
0.16--0.18& 17.10&$\pm$&  0.03&$\pm$&  0.16\\
0.18--0.20& 13.54&$\pm$&  0.03&$\pm$&  0.14\\
0.20--0.25&  9.37&$\pm$&  0.01&$\pm$&  0.11\\
0.25--0.30&  5.66&$\pm$&  0.01&$\pm$&  0.08\\
0.30--0.40&  2.89&$\pm$&  0.01&$\pm$&  0.05\\
0.40--0.50&  1.208&$\pm$&  0.004&$\pm$&  0.036\\
0.50--0.60&  0.506&$\pm$&  0.002&$\pm$&  0.018\\
0.60--0.80&  0.153&$\pm$&  0.001&$\pm$&  0.012\\
0.80--1.00&  0.0199&$\pm$&  0.0003&$\pm$&  0.0044\\

%
%
                                           \hline
\end{tabular}
\caption{\sl Inclusive fragmentation function. The first error is statistical, the second 
systematic.
\label{Table_tot}}
\end{center}
\end{table} 
%
%
\begin{table}[p]
\begin{center}
\begin{tabular}{|c|c@{}c@{}c@{}c@{}c|c@{} c@{} c@{} c@{} c|c@{}  c @{}c@{} c@{} c|}   
                            \hline
   \xip   &  \multicolumn{15}{|c|}
       {$1/{\sigma_{\mathrm{tot}}}\cdot{\deriv{\sigma^h}}/{\deriv{\xip}}$}  \\
       \cline{2-16}
   & \multicolumn{5}{|c|}{uds events}  
   & \multicolumn{5}{|c|}{c events} 
   & \multicolumn{5}{|c|}{b events}\\
                            \hline
                            \hline
%
%
0.0--0.2
&$  0.024$&$\pm$&$  0.006$&$\pm$&$  0.006$
&$  0.002$&$\pm$&$  0.147$&$\pm$&$  0.005$
&$  0.0034$&$\pm$&$  0.0001$&$\pm$&$  0.0027$
\\
0.2--0.4
&$  0.114$&$\pm$&$  0.003$&$\pm$&$  0.011$
&$  0.005$&$\pm$&$  0.123$&$\pm$&$  0.008$
&$  0.014$&$\pm$&$  0.001$&$\pm$&$  0.010$
\\
0.4--0.6
&$  0.277$&$\pm$&$  0.009$&$\pm$&$  0.025$
&$  0.07$&$\pm$&$  0.03$&$\pm$&$  0.06$
&$  0.066$&$\pm$&$  0.002$&$\pm$&$  0.023$
\\
0.6--0.8
&$  0.529$&$\pm$&$  0.016$&$\pm$&$  0.032$
&$  0.20$&$\pm$&$  0.06$&$\pm$&$  0.07$
&$  0.188$&$\pm$&$  0.004$&$\pm$&$  0.034$
\\
0.8--1.0
&$  0.86$&$\pm$&$  0.02$&$\pm$&$  0.05$
&$  0.52$&$\pm$&$  0.08$&$\pm$&$  0.11$
&$  0.40$&$\pm$&$  0.01$&$\pm$&$  0.05$
\\
1.0--1.2
&$  1.31$&$\pm$&$  0.03$&$\pm$&$  0.06$
&$  0.84$&$\pm$&$  0.12$&$\pm$&$  0.12$
&$  0.71$&$\pm$&$  0.01$&$\pm$&$  0.07$
\\
1.2--1.4
&$   1.76$&$\pm$&$   0.05$&$\pm$&$   0.07$
&$   1.43$&$\pm$&$   0.16$&$\pm$&$   0.15$
&$   1.15$&$\pm$&$   0.01$&$\pm$&$   0.07$
\\
1.4--1.6
&$   2.22$&$\pm$&$   0.06$&$\pm$&$   0.06$
&$   2.10$&$\pm$&$   0.20$&$\pm$&$   0.14$
&$   1.74$&$\pm$&$   0.01$&$\pm$&$   0.09$
\\
1.6--1.8
&$   2.70$&$\pm$&$   0.07$&$\pm$&$   0.06$
&$   2.88$&$\pm$&$   0.24$&$\pm$&$   0.18$
&$   2.49$&$\pm$&$   0.02$&$\pm$&$   0.10$
\\
1.8--2.0
&$   3.06$&$\pm$&$   0.08$&$\pm$&$   0.09$
&$   3.89$&$\pm$&$   0.29$&$\pm$&$   0.29$
&$   3.41$&$\pm$&$   0.02$&$\pm$&$   0.08$
\\
2.0--2.2
&$   3.76$&$\pm$&$   0.09$&$\pm$&$   0.11$
&$   3.70$&$\pm$&$   0.31$&$\pm$&$   0.34$
&$   4.30$&$\pm$&$   0.02$&$\pm$&$   0.13$
\\
2.2--2.4
&$   4.03$&$\pm$&$   0.10$&$\pm$&$   0.13$
&$   4.77$&$\pm$&$   0.35$&$\pm$&$   0.31$
&$   5.19$&$\pm$&$   0.02$&$\pm$&$   0.11$
\\
2.4--2.6
&$   4.48$&$\pm$&$   0.10$&$\pm$&$   0.18$
&$   5.32$&$\pm$&$   0.37$&$\pm$&$   0.45$
&$   5.98$&$\pm$&$   0.02$&$\pm$&$   0.15$
\\
2.6--2.8
&$   5.12$&$\pm$&$   0.11$&$\pm$&$   0.16$
&$   4.83$&$\pm$&$   0.37$&$\pm$&$   0.21$
&$   6.46$&$\pm$&$   0.03$&$\pm$&$   0.15$
\\
2.8--3.0
&$   5.22$&$\pm$&$   0.12$&$\pm$&$   0.17$
&$   6.39$&$\pm$&$   0.41$&$\pm$&$   0.31$
&$   6.92$&$\pm$&$   0.03$&$\pm$&$   0.15$
\\
3.0--3.2
&$   5.26$&$\pm$&$   0.13$&$\pm$&$   0.19$
&$   7.89$&$\pm$&$   0.48$&$\pm$&$   0.37$
&$   7.19$&$\pm$&$   0.03$&$\pm$&$   0.16$
\\
3.2--3.4
&$   6.24$&$\pm$&$   0.12$&$\pm$&$   0.21$
&$   5.48$&$\pm$&$   0.43$&$\pm$&$   0.40$
&$   7.37$&$\pm$&$   0.03$&$\pm$&$   0.17$
\\
3.4--3.6
&$   6.02$&$\pm$&$   0.12$&$\pm$&$   0.20$
&$   6.90$&$\pm$&$   0.44$&$\pm$&$   0.36$
&$   7.49$&$\pm$&$   0.03$&$\pm$&$   0.17$
\\
3.6--3.8
&$   5.89$&$\pm$&$   0.13$&$\pm$&$   0.26$
&$   7.12$&$\pm$&$   0.45$&$\pm$&$   0.74$
&$   7.49$&$\pm$&$   0.03$&$\pm$&$   0.16$
\\
3.8--4.0
&$   6.04$&$\pm$&$   0.12$&$\pm$&$   0.20$
&$   6.50$&$\pm$&$   0.43$&$\pm$&$   0.40$
&$   7.26$&$\pm$&$   0.03$&$\pm$&$   0.16$
\\
4.0--4.2
&$   5.85$&$\pm$&$   0.13$&$\pm$&$   0.20$
&$   6.24$&$\pm$&$   0.46$&$\pm$&$   0.49$
&$   6.93$&$\pm$&$   0.03$&$\pm$&$   0.14$
\\
4.2--4.4
&$   5.58$&$\pm$&$   0.11$&$\pm$&$   0.14$
&$   5.67$&$\pm$&$   0.40$&$\pm$&$   0.35$
&$   6.32$&$\pm$&$   0.03$&$\pm$&$   0.09$
\\
4.4--4.6
&$   5.15$&$\pm$&$   0.11$&$\pm$&$   0.09$
&$   4.84$&$\pm$&$   0.40$&$\pm$&$   0.16$
&$   5.76$&$\pm$&$   0.03$&$\pm$&$   0.08$
\\
4.6--4.8
&$   4.21$&$\pm$&$   0.12$&$\pm$&$   0.24$
&$   5.76$&$\pm$&$   0.41$&$\pm$&$   0.80$
&$   4.91$&$\pm$&$   0.02$&$\pm$&$   0.05$
\\
4.8--5.0
&$   3.99$&$\pm$&$   0.10$&$\pm$&$   0.14$
&$   3.35$&$\pm$&$   0.36$&$\pm$&$   0.43$
&$   4.24$&$\pm$&$   0.02$&$\pm$&$   0.07$
\\
5.0--5.2
&$   2.94$&$\pm$&$   0.10$&$\pm$&$   0.15$
&$   3.34$&$\pm$&$   0.33$&$\pm$&$   0.45$
&$   3.30$&$\pm$&$   0.02$&$\pm$&$   0.06$
\\
5.2--5.4
&$   2.14$&$\pm$&$   0.10$&$\pm$&$   0.12$
&$   3.22$&$\pm$&$   0.34$&$\pm$&$   0.40$
&$   2.54$&$\pm$&$   0.02$&$\pm$&$   0.06$
\\
5.4--5.6
&$   1.93$&$\pm$&$   0.08$&$\pm$&$   0.13$
&$   1.36$&$\pm$&$   0.28$&$\pm$&$   0.32$
&$   1.92$&$\pm$&$   0.02$&$\pm$&$   0.09$
\\
5.6--5.8
&$   1.43$&$\pm$&$   0.09$&$\pm$&$   0.23$
&$   0.78$&$\pm$&$   0.32$&$\pm$&$   0.62$
&$   1.42$&$\pm$&$   0.03$&$\pm$&$   0.16$
\\

%
%
                                           \hline
\end{tabular}
\caption{\sl $\xip\ = \ln(1/\xe)$ distribution of uds, c and b events. The first
error is statistical, the second 
systematic.
\label{Table_xi_uds}}
\end{center}
\end{table}
\begin{table}[p]
\begin{center}
\begin{tabular}{|c|c@{ }c@{ }c@{ }c@{ }c|}   
                            \hline
   \xip   &  \multicolumn{5}{|c|}
       {$1/{\sigma_{\mathrm{tot}}}\cdot{\deriv{\sigma^h}}/{\deriv{\xip}}$}  \\
                            \hline
                            \hline
%
%
0.0--0.2
&$  0.0153$&$\pm$&$  0.0003$&$\pm$&$  0.0035$
\\
0.2--0.4
&$  0.071$&$\pm$&$  0.001$&$\pm$&$  0.006$
\\
0.4--0.6
&$  0.191$&$\pm$&$  0.001$&$\pm$&$  0.008$
\\
0.6--0.8
&$  0.392$&$\pm$&$  0.001$&$\pm$&$  0.011$
\\
0.8--1.0
&$  0.693$&$\pm$&$  0.002$&$\pm$&$  0.018$
\\
1.0--1.2
&$  1.087$&$\pm$&$  0.002$&$\pm$&$  0.021$
\\
1.2--1.4
&$  1.556$&$\pm$&$  0.003$&$\pm$&$  0.025$
\\
1.4--1.6
&$  2.082$&$\pm$&$  0.003$&$\pm$&$  0.028$
\\
1.6--1.8
&$  2.674$&$\pm$&$  0.004$&$\pm$&$  0.028$
\\
1.8--2.0
&$  3.272$&$\pm$&$  0.004$&$\pm$&$  0.032$
\\
2.0--2.2
&$  3.866$&$\pm$&$  0.005$&$\pm$&$  0.036$
\\
2.2--2.4
&$  4.403$&$\pm$&$  0.005$&$\pm$&$  0.044$
\\
2.4--2.6
&$  4.96$&$\pm$&$  0.01$&$\pm$&$  0.06$
\\
2.6--2.8
&$  5.39$&$\pm$&$  0.01$&$\pm$&$  0.07$
\\
2.8--3.0
&$  5.82$&$\pm$&$  0.01$&$\pm$&$  0.08$
\\
3.0--3.2
&$  6.15$&$\pm$&$  0.01$&$\pm$&$  0.10$
\\
3.2--3.4
&$  6.38$&$\pm$&$  0.01$&$\pm$&$  0.11$
\\
3.4--3.6
&$  6.52$&$\pm$&$  0.01$&$\pm$&$  0.12$
\\
3.6--3.8
&$  6.48$&$\pm$&$  0.01$&$\pm$&$  0.10$
\\
3.8--4.0
&$  6.41$&$\pm$&$  0.01$&$\pm$&$  0.12$
\\
4.0--4.2
&$  6.17$&$\pm$&$  0.01$&$\pm$&$  0.09$
\\
4.2--4.4
&$  5.75$&$\pm$&$  0.01$&$\pm$&$  0.06$
\\
4.4--4.6
&$  5.24$&$\pm$&$  0.01$&$\pm$&$  0.04$
\\
4.6--4.8
&$  4.65$&$\pm$&$  0.01$&$\pm$&$  0.04$
\\
4.8--5.0
&$  3.95$&$\pm$&$  0.01$&$\pm$&$  0.05$
\\
5.0--5.2
&$  3.110$&$\pm$&$  0.005$&$\pm$&$  0.051$
\\
5.2--5.4
&$  2.423$&$\pm$&$  0.004$&$\pm$&$  0.055$
\\
5.4--5.6
&$  1.835$&$\pm$&$  0.004$&$\pm$&$  0.081$
\\
5.6--5.8
&$  1.33$&$\pm$&$  0.01$&$\pm$&$  0.15$
\\

%
%
                                           \hline
\end{tabular}
\caption{\sl Inclusive $\xip\ = \ln(1/\xe)$ distribution. The first error is statistical, the second 
systematic.
\label{Table_xi_tot}}
\end{center}
\end{table} 
%
%
\newpage
\begin{table}[hbt]
\begin{center}
\begin{tabular}{|l | c c|c c|c c|c c|}  
                            \hline 
    Effect & \multicolumn{2}{| c |}{ inclusive } &  \multicolumn{2}{| c |}{ uds } &
             \multicolumn{2}{| c |}{  c  } &  \multicolumn{2}{| c |}{  b  }\\
                            \hline      
  (1) Purities of sec.~vertex samples &  - & ( - )  &  2.2 & ( 4.0)  &  0.9 & ( 1.5)  &  1.4 & ( 6.5) \\
        (2) Purities of \Dstar\ samples &  - & ( - )  &  0.2 & ( 0.3)  &  0.5 & ( 1.0) 
	&  $<0.1$ & ($<0.1$) \\
              (3) \Dstar\ selection &  - & ( - )  &  1.2 & ( 1.8)  &  2.6 & ( 4.2)  &  $<0.1$ & ( 0.1) \\
         (4) Hemisphere correlation &  - & ( - )  &  0.2 & ( 0.9)  &  0.4 & ( 0.6)  &  0.3 & ( 0.1) \\
           (5) Correction procedure &  $<0.1$ & ($<0.1$)  &  0.8 & ( 1.7)  &  1.0 & ( 3.3)  &  1.7 & ( 7.6) \\
      (6) Track and event selection &  1.5 & ( 1.9)  &  1.5 & ( 1.9)  &  1.5 & ( 1.9)  &  1.5 & ( 1.9) \\
 \hline
     Total systematic uncertainty &  1.5 & ( 1.9)  &  3.0 & ( 5.2)  &  3.4 & ( 6.0)  &  2.6 & (10.2) \\
 \hline
      Total statistical uncertainty &  0.1 & ( 0.2)  &  2.3 & ( 2.6)  &  6.7 & (12.5)  &  0.4 & ( 1.0) \\
 \hline

\end{tabular}
\caption {\sl Relative systematic and statistical uncertainties in percent 
on the results for $0.05 < \xe < 0.06$ ($0.3 <\xe < 0.4$).
 \label{tabsys1}}
\end{center}
\end{table} 
%
%
\begin{table}[hbt]
\begin{center}
\begin{tabular}{|l | c |c |c |c |}  
                            \hline 
    Effect & inclusive  & uds  & c   &  b \\
                            \hline      
  (1) Purities of sec.~vertex samples &  -   &  1.4   &  0.8   &  1.3  \\
        (2) Purities of \Dstar\ samples &  -   &  0.1   &  0.3   &  $<0.1$  \\
              (3) \Dstar\ selection &  -   &  0.6   &  2.1   &  $<0.1$  \\
         (4) Hemisphere correlation &  -   &  $<0.1$   &  0.1   &  0.2  \\
(5) Correction procedure &  0.04   &  0.1   &  1.3   &  1.7  \\
      (6) Track and event selection &  0.96   &  1.0   &  1.0   &  1.0  \\
 \hline
     Total systematic uncertainty &  0.96   &  1.8   &  2.8   &  2.4  \\
 \hline
      Total statistical uncertainty &  0.03   &  0.6   &  2.0   &  0.1  \\
 \hline

\end{tabular}
\caption {\sl Relative systematic and statistical uncertainties in percent 
on the results for the mean value of the \xe\ distribution \xemean.
 \label{tabsys2}}
\end{center}
\end{table} 
%
%
%
\begin{table}[hbt]
\begin{center}
\begin{tabular}{|l | c |c |c |c |}  
                            \hline 
    Effect & inclusive  & uds  & c   &  b \\
                            \hline      
  (1) Purities of sec.~vertex samples &  -   &  0.6   &  0.1   &  0.2  \\
       (2) Purities of \Dstar\ samples &  -   &  $<0.1$   &  0.2   &  $<0.1$  \\
               (3) \Dstar\ selection &  -   &  0.5   &  2.0   &  $<0.1$  \\
         (4) Hemisphere correlation &  -   &  0.4   &  0.2   &  $<0.1$  \\
           (5) Correction procedure &  0.9   &  0.6   &  0.3   &  0.5  \\
      (6) Track and event selection &  0.6   &  0.6   &  0.6   &  0.6  \\
         (7) Fit range and fit type &  3.0   &  5.3   &  8.4   &  1.9  \\
 \hline
       Total systematic uncertainty &  3.1   &  5.4   &  8.7   &  2.1  \\
 \hline
      Total statistical uncertainty &  0.1   &  1.5   &  4.1   &  0.3  \\
 \hline

\end{tabular}
\caption {\sl Relative systematic and statistical uncertainties in percent 
on the results for the position of the maximum of the \xip\ distribution, \ximax.
 \label{tabsys3}}
\end{center}
\end{table} 
%
%
%
%
%
%
%
%
%
%
%
%
%
%
%
%
\end{document}